\documentstyle[aaspp4]{article}

\def \etal{et~al.\/}

\begin{document}

%
%

\title{SED Signatures of Jovian Planets Around White Dwarf Stars
        }
\author{R.~Ignace\altaffilmark{1,}\altaffilmark{2}}

\altaffiltext{1}{
        Email:  ri@astro.physics.uiowa.edu }
 
\altaffiltext{2}{
        203 Van Allen Hall,
        Department of Physics and Astronomy,
        University of Iowa,
        Iowa City, IA 52242 USA }

\begin{abstract}

The problem of detecting Jovian-sized planets orbiting White Dwarf
stars is considered.  Significant IR excesses result from warm Jupiters
orbiting a White Dwarf of $T_{\rm eff}=10000$~K at a distance of $\sim
10^3$ White Dwarf radii (corresponding to $\sim 10^2$ Jupiter radii or
a few tenths of an AU) with an orbital period of $\sim 100$~days.  Such
a planet will have a 10 micron flux density at its Wien peak that is
comparable to the emission of the White Dwarf at that wavelength.
Although the White Dwarf is much hotter than the planet, the planet
will have peak brightness at the IR, well into the Rayleigh-Jeans tail
of the White Dwarf, plus Jovians are about 10 times larger than White
Dwarfs, so there is a substantial gain in the planet to star brightness
contrast as compared to planets around Main Sequence stars.  In the
solar neighborhood, there are 51 White Dwarf stars within 13~pc of the
Sun.  At 10~pc, the IR flux density of ``warm'' Jupiters (a few hundred
Kelvin) will fall in the range 10--100 micro-Jansky which should be
observable with {\it SIRTF}.

\end{abstract}


\section{Introduction}

There is tremendous excitement in the astronomical community regarding
recent discoveries of extrasolar planets and future projects for
accelerating detection rates and expanding the range of planetary types
that can be observed (e.g., see reviews by Marcy \& Butler
1998; Woolf \& Angel 1998).  The method which has been so successful in
the ongoing discovery of extra-solar planets is based on Doppler shift
observations of stellar photospheric lines, an indirect method for
inferring the presence of an orbiting companion (e.g., Mayor
\& Queloz 1995; Marcy \& Butler 1996; Butler \& Marcy 1996).  Other
methods have been used in the search for extra-solar planets but
without as much success as the Doppler technique.  For example, direct
detection through imaging is hampered both by the faintness of planets
and the brightness of the central stars.  Transit events are similarly
difficult to detect owing to the small brightness variations of the
star and the requirement for a special viewing orientation; however,
the transit of HD~209458 by its planetary companion demonstrates the
tremendous potential for gleaning important information about a planet
during such occurrences (Charbonneau \etal\ 2000; Henry \etal\ 2000;
Brown \etal\ 2001).

Some have considered the possibility of detecting planets around White
Dwarf stars.  The problem has been approached from several different
directions.  For example, Provencal (1997) has discussed the
possibility of phase analysis techniques for pulsating White Dwarfs.
Li, Ferrario, \& Wickramasinghe (1998) consider methods based on the
interaction of Jovian planets with the magnetosphere of the White Dwarf
(in analogy to the Jupiter-Io system).  Chu \etal\ (2001) discuss a
technique based on hydrogen recombination lines formed in the Jovian
atmosphere for planets around UV-bright White Dwarfs.  Livio, Pringle,
\& Saffer (1992) comment that an Earth-like planet with a temperature
of 300~K would produce an IR excess of about 1\% of the White Dwarf
continuum emission.  The contribution of this paper is to consider the
IR~excess for Jovian-type planets at a range of orbital radii, and to
make quantitative estimates on the expected continuum emission and the
number of nearby targets that could reasonably be expected to yield
possible detections.

Observationally, the major advantage of searching for giant planets
around White Dwarfs is that they are much smaller in diameter than
solar-type Main Sequence stars.  They also tend to be somewhat hotter
than G~dwarfs.  Using the 4$^{th}$ editionof the Catalog of
Spectroscopically Idenfitifed White Dwarfs (McCook \& Sion 1999),
Holberg, Oswalt, \& Sion (2001) have identified the known White Dwarf
stars within 20 pcs.  Using the type designation provided in their
Table~1 to estimate effective surface temperatures, I find an average
value of about 8700~K for 103 stars (approximately one-fourth of the
stars have temperatures in excess of 10000~K, up to about 25000~K).  In
the net, White Dwarfs tend to have luminosities of about a thousand
times smaller than solar-type stars.  So there is potentially a strong
gain in the brightness contrast of a planet around a White Dwarf star
when compared to a late-type Main Sequence star.

However, it is not quite so straightforward.  Consider the elementary
expression for the temperature $T_P$ of a planet with an albedo, $A_P$,
in a circular orbit of radius $d_P$ about a White Dwarf of radius
$R_{WD}$ and temperature $T_{WD}$:

\begin{equation}
T_P = T_{WD} \, (1-A_P)^{1/4} \, \left(\frac{R_{WD}}{2d_P}\right)^{1/2}.
	\label{eq:tp_elem}
\end{equation}

\noindent The ratio of Bolometric luminosities between the planet and
the White Dwarf will be

\begin{equation}
\frac{L_P}{L_{WD}} = \frac{R_P^2\,T_P^4}{R_{WD}^2\,T_{WD}^4} =
	\left(\frac{R_P}{2d_P}\right)^2\,(1-A_P),
\end{equation}

\noindent so that the {\it Bolometric} brightness contrast between the
planet and star does not actually depend on the star at all, only the
radius of the planet, its distance from the star, and the albedo.  (Of
course, cooler stars will be fainter, and all things being equal, their
planets will be cooler and fainter too, so that longer integration
times will be needed to reach a given detection threshold.) On the
other hand, the orbits of planets can drastically change during a
star's post-Main Sequence evolution, such as might occur during a
common envelope phase (e.g., Livio \& Soker 1984).  It may be possible
that $d_P$ could degrade to small values.  In some cases a low-mass
companion will spiral into the star during its bloated AGB~phase; in
other cases a companion may drift to orbits of larger size.  Given that
the discovery of large planets orbiting solar-type stars with periods
of a few days came as a great surprise to the astronomical community,
nature may have a way of placing Jovian planets in orbits of small or
modest size ($\lesssim 1$~AU) around White Dwarf stars.  Taking the
purely empirical approach, what are the observational signatures that
should be considered to determine in fact whether planets around
low-mass stellar remnants are common or rare?

The approach adopted here will be to consider spectroscopic signatures
in terms of the continuum energy distribution from a White Dwarf $+$
Jovian Planet system.  These results should also be applicable to
binary systems involving a White Dwarf and a Brown Dwarf (e.g., as
discussed by Stringfellow, Black, \& Bodenheimer 1990).  If the system
is a close binary, the Brown Dwarf companion might not be resolvable as
a secondary source but could betray its presence through an IR~excess.

The following section details the flux density estimates for such a
planet orbiting a White Dwarf in terms of the central star's
temperature and the planet's orbital proximity.  The paper concludes
with a discussion of the observational prospects.

\section{The Combined Spectrum of a White Dwarf and Jovian Planet}

To compute the spectral energy distribution (SED) of a White Dwarf and
a planetary companion, it shall be assumed that both objects emit a
blackbody spectrum.  Specifying the temperature and radius of the White
Dwarf allows one to calculate the planet's temperature assuming
radiative equilibrium and provided values for the albedo, planet
radius, and orbital distance.  However in considering a planet that
orbits quite close to the White Dwarf, one cannot use the elementary
form for computing the planet temperature.  Also, one must realize that
a Jovian planet is an order of magnitude larger than the White Dwarf.
This means that the shortest orbital distance between the two is $d_P =
R_{WD} + R_P \approx R_P$.  Actually, this is a geometric limit.  A
more fundamental physical barrier is the Roche limit.  For a typical
White Dwarf and a Jupiter-like planet, the classical Roche limit (in
this case $r_{\rm clas} = 2.44R_{WD}\,(\rho_{WD}/\rho_P)^{1/3}$ for a
liquid spherical planet) yields a value of 270~$R_{WD}$, or 24~$R_J$.
However, continuing with an exact derivation of the planet temperature
yields

\begin{equation}
T_P = T_{WD}\,(1-A_P)^{1/4}\,\left(\frac{R_{WD}}{R_P}\right)^{1/2}\,
	[W(R_P,d_P)]^{1/4} ,
	\label{eq:tp}
\end{equation}

\noindent where the dilution factor $W$ is given by

\begin{equation}
W(R_P,d_P) = \frac{1}{2}\,\left(1-\sqrt{1-\frac{R_P^2}{d_P^2}}\,\right).
\end{equation}

\noindent In the limit that $(R_P/d_P)^2 \ll 1$, the dilution factor
approximates to $W\approx 0.25\,(R_P/d_P)^2$, and
equation~(\ref{eq:tp}) reduces to the standard elementary form.

\begin{figure}
\plotone{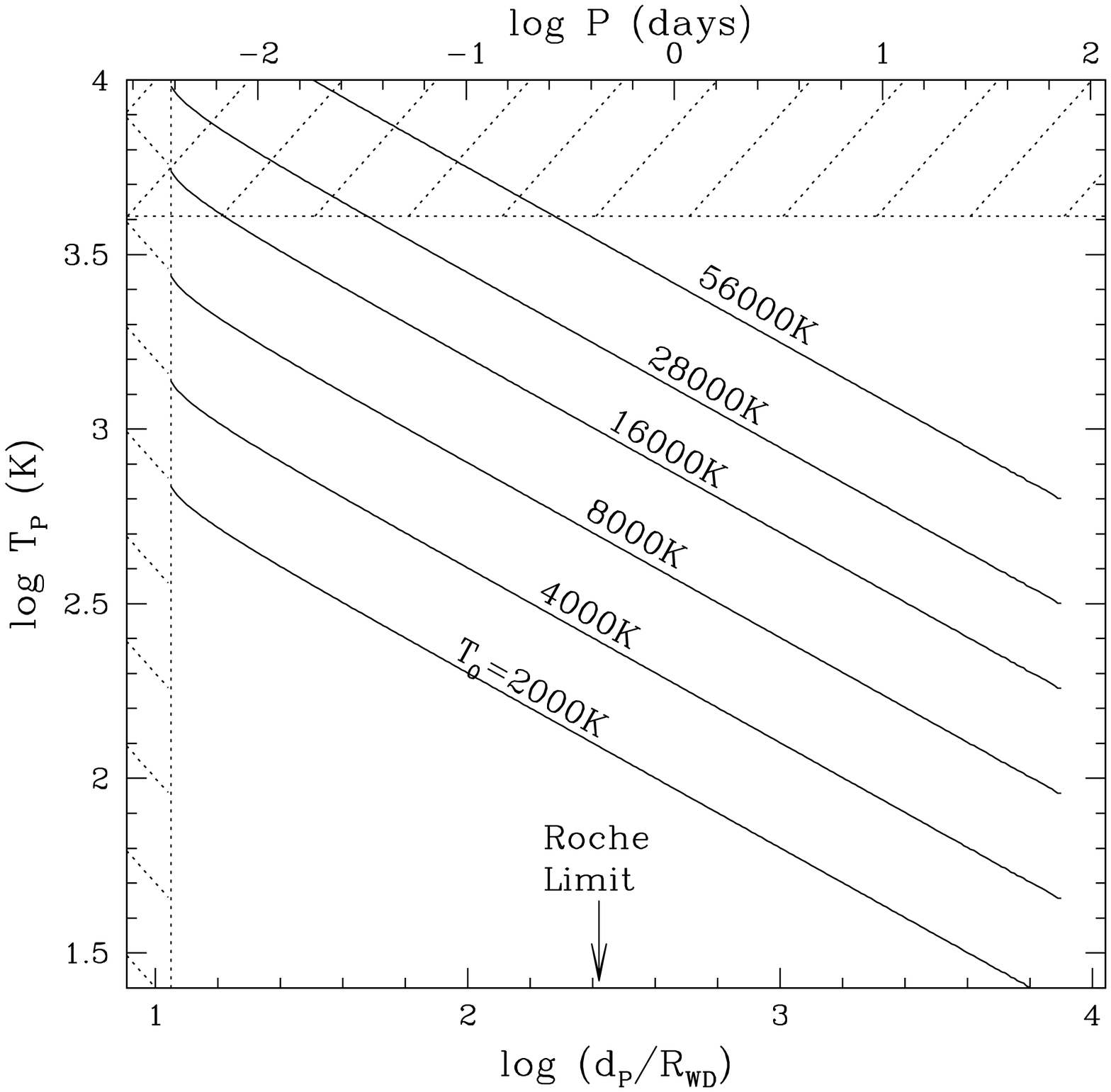}
\caption[]{A plot of the expected planetary temperature $T_P$ for
Jupiter-sized gas giant in a circular orbit of radius $d_P$ from a
White Dwarf star of radius $R_{WD}=R_\oplus$ and mass
$M_{WD}=1M_\odot$.  The upper axis indicates the orbital period $P$ in
days.  The five curves are distinguished by the temperature parameter
$T_0 = 2^{-1/2}\,T_{WD}\,(1-A_P)^{1/4}$ as indicated.  (For $A_P=0.5$,
the curves correspond to $T_{WD} \approx 3400$, 6700, 13500, 27000,
47000, and 94000~K.) The shaded region at top is a ``forbidden zone''
where a Jovian would obtain to such a high value of $T_P$ that hydrogen
would readily escape from the planet.  At left is a forbidden region
corresponding to $d_P=R_{WD}+R_J$, although in reality a Jovian would
never get this close since the Roche limit falls at around $d_P \approx
270 R_{WD}$.

\label{fig:f1}} \end{figure}

Figure~\ref{fig:f1} plots curves of $T_P$ against the orbital distance
$d_P$.  The lower scale is for $d_P/R_{WD}$, and the upper scale is for
the orbital period $P$ in days, assuming that $R_{WD} = R_\oplus$.
Each curve is for a different temperature parameter $T_0$ associated
with the White Dwarf.  This parameter is defined by

\begin{equation}
T_0 = 2^{-1/2}\,(1-A_P)^{1/4}\,T_{WD},
\end{equation}

\noindent which is essentially the constant coefficient for the
elementary form of $T_P$ if only $d_P/R_{WD}$ is allowed to vary.
From lower left to upper right, the curves are for $T_0 = 2000$, 4000,
8000, 16000, 28000, and 56000~K.  For comparison, Chu \etal\ (2001)
considered models with $T_{WD}$ in the range of 20000--200000~K for
$d_P$ between 0.5 and 5~AU ($10^4-10^5 R_{WD}$) for gas giants of $R_P$
between 0.5 and 5.0~$R_J$.  Assuming $A_P\lesssim 0.5$, the curves in
Figure~\ref{fig:f1} have some overlap with the temperature range
considered by Chu \etal, but are generally for cooler stars as
appropriate for the sample of nearby White Dwarfs; however, those
authors targeted hot White Dwarfs, which although fairly rare, provide
the requisite UV flux for the planetary recombination line analysis
that they were investigating.

Assuming that the planet has a radius like that of Jupiter, the Figure
shows an excluded region at left corresponding to $d_P <
(R_J+R_{WD})$.  The arrow at bottom indicates the classical Roche
limit.  There is also a forbidden region at top, where the planet
becomes so hot that hydrogen gas can readily escape the planet's
atmosphere.  This is set by the condition that $v_{\rm th}({\rm H})
\gtrsim \slantfrac{1}{6} v_{\rm esc}$, corresponding to $T_P \gtrsim
4100$~K.

For a particular set of values for the White Dwarf and planet
parameters, the specific fluxes are given respectively by

\begin{equation}
F_\nu^{WD} = \frac{\pi\,R_{WD}^2}{4\,D^2}\,B_\nu (T_{WD}) =(3.4\times 
	10^6\,{\rm mJy})\times \frac{R_\oplus^2}{D^2_{\rm pc}}\,B_\nu (T_{WD}),
\end{equation}

\noindent and

\begin{equation}
F_\nu^{P} = \frac{\pi\,R_{P}^2}{4\,D^2}\,B_\nu (T_{P}) = (4.2\times 
        10^8\,{\rm mJy})\times \frac{R_{J}^2}{D^2_{\rm pc}}\,B_\nu (T_{P}),
\end{equation}

\noindent where $R_\oplus$ signifies the White Dwarf radius in Earth
radii, $R_J$ signifies the planet radius in Jupiter radii, $D_{\rm pc}$
indicates the distance to the system in parsecs, and $B_\nu$ must be
given in cgs units.  Figure~\ref{fig:f2} shows SEDs for a White Dwarf
at $D=1$~pc with fixed values of $T_{WD}=10000$~K and $R_{WD}=R_\oplus$
but Jovian sized planets of various temperatures.  The frequency range
spans from the UV out to the sub-mm.  The dotted line is for the White
Dwarf, the dashed lines for the planet, and the solid lines represent
the combined spectra.  The three planetary curves are for $T_P = 150$,
225, and 450~K (this latter value is the approximately maximum possible
based on the Jovian being at the classical Roche limit).  The SEDs can
be significantly modified from a Planckian for hot Jupiters in the IR
(even the NIR for planets around hotter White Dwarfs).  For cool
Jupiters, the long wavelength spectral slope deviates from the power
law of $\nu^2$ for the Rayleigh-Jeans tail in the vicinity of the Wien
peak for the planet.  For warm Jupiters of several hundred Kelvin, the
IR excess occurs at the level of tens of milli-Jansky.  At sufficiently
long wavelengths, both the planet and the White Dwarf SED will vary as
$\nu^2$; however, the presence of the planet can still be inferred (for
a sufficiently sensitive instrument) from the excess emission.  In the
Rayleigh-Jeans portion for the combined spectrum, the ratio of total
emission to that of just the White Dwarf will be given by $1 + (R_P^2
T_P)/(R_{WD}^2 T_{WD})\approx 1 + 100 T_P/T_{WD}$, with values ranging
typically from 1 (planet not detectable) to 10 (long wavelength
emission dominated by the planet).  It has been assumed that for the
analysis of an observed spectrum, the value of $T_{WD}$ is already
known, and that other sources of long wavelength emission are
incommensurate with the continuum slope or amount of emission.

\section{Discussion}

White Dwarfs are interesting as candidates for harboring planets
because (a) planets are thought to be a common occurrence around
solar-type stars, and White Dwarfs are the endstates of low-mass
stellar evolution, (b) it is possible for planets to suffer orbital
degradation during evolved stages of evolution, and (c) White Dwarfs
are fairly common ($\sim 10^{-2}$ pc$^{-3}$) and thus there are a
significant number of them to be found in the locale of the Sun.  This
paper reports on the possibility of detecting extra-solar planets
around White Dwarf stars through an analysis of SEDs.  The Roche limit
is found to be an important limiting factor on the temperature and
hence brightness of planets.  Using the classical limit, the planet
cannot orbit any closer than about 24~$R_J$, and so its temperature is
limited to $T_P \lesssim 0.043 T_{WD}$.  This means that the typical
nearby White Dwarfs could have warm Jupiters, but not hot ones.  The
primary observational diagnostics arise from (a) an IR ``bump'', or
softening of the power law spectral index, in the SED around the
vicinity of the Wien peak for the planet but well into the
Rayleigh-Jeans tail of the White Dwarf, or (b) significant excess
emission in the Rayleigh-Jeans tail of the combined emission.

\begin{figure}
\plotone{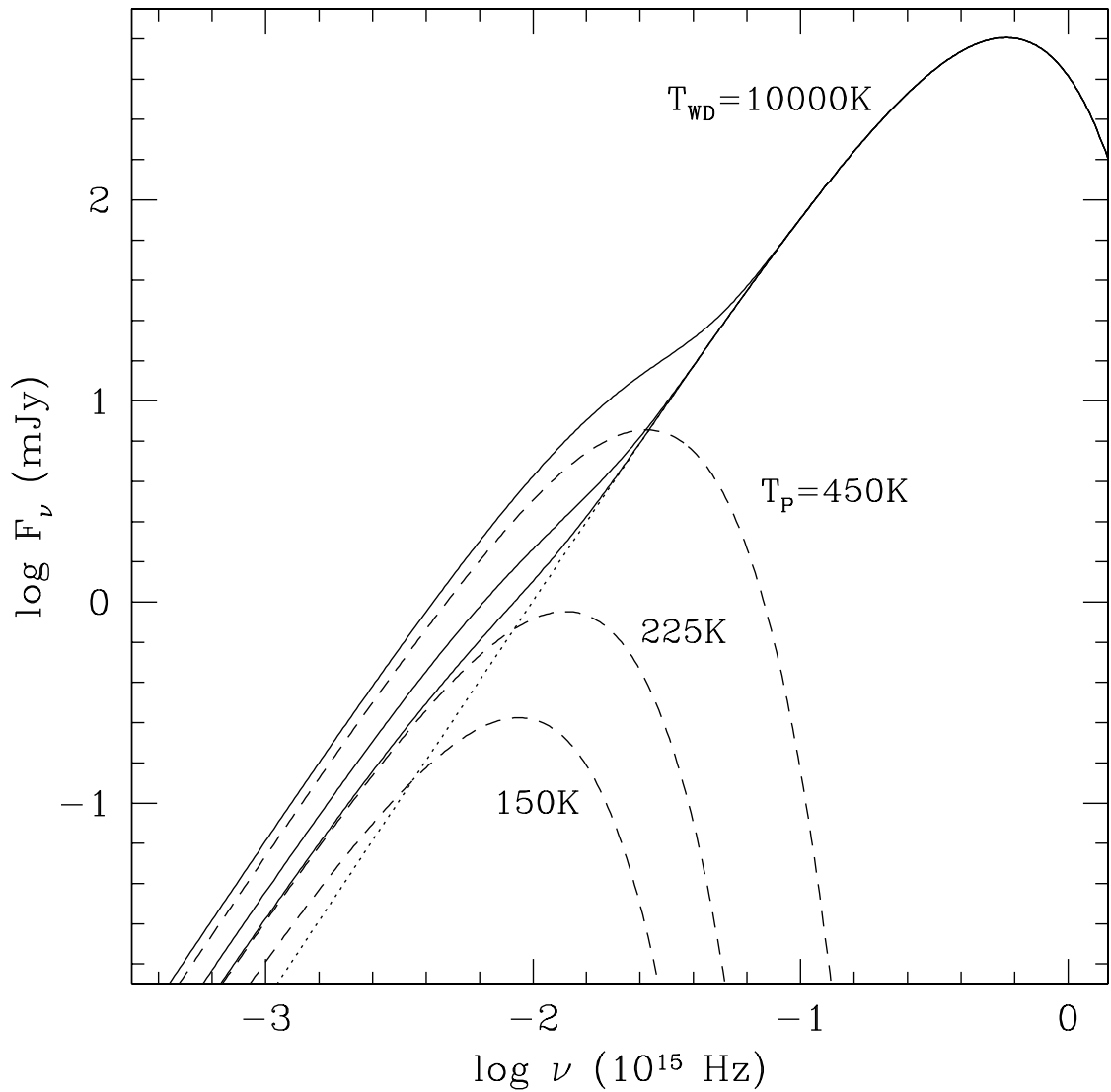}
\caption[]{Spectral energy distributions in milli-Janskys for a 10000~K
White Dwarf (dotted line), four different planetary temperatures
(dashed), and the combined flux densities (solid).  The planetary SEDs
are calculated for $T_P=150$, 225, and 450~K.  Although asymptotically,
the low frequency slope of the SEDs is that of the Rayleigh-Jeans tail,
the presence of a planet can produce a strong IR/sub-mm excess above
the White Dwarf spectrum, and in the vicinity of the Wien peak for each
planet, the merged SED significantly deviates from Planckian.  Note
that the flux density scale as shown applies for a White Dwarf $+$
Jupiter system at $D=1$~pc.

\label{fig:f2}}
\end{figure}

In the tabulation of nearby White Dwarf stars, Holberg \etal\ (2001)
determine that the catalog is complete to 13 pcs, and that there are 51
White Dwarfs within a spherical volume of that radius (from their
Tab.~1).  Of these, 17 are binary systems (two are double-degenerate
systems), leaving 32 single star candidates to target for a search of
planetary companions.  The SEDs in Figure~\ref{fig:f2} were computed
for a White Dwarf at 1~pc.  A ``warm'' Jupiter of a few hundred Kelvin
orbiting a White Dwarf at a distance of 10~pc should have a flux
density at its Wien peak (10--30 microns) that is comparable to that of
a 10000~K White Dwarf with a value in the range of 10--100
micro-Janskys.  Such faint signals should be detectable, for example,
with SIRTF.  The IRAC camera conducts IR photometry at 3.6, 4.5, 5.8,
and 8.0 microns (with bandpasses of 1--2 microns in width).  At
8.0~microns, the IRAC can detect an 18 micro-Jansky source at $5\sigma$
in about 8 minutes (less time is required at the shorter wavelengths).
This would be sufficient to determine whether or not a White Dwarf
exhibited an IR excess.  For the same time and S/N, the low resolution
short wavelength spectrograph on the IRS could build up a spectrum for
a 550 micro-Jansky source in the range 5--14 microns.  (Observing time
estimates for both IRAC and IRS are based on information provided by
the SIRTF Science Center at the web site
http://sirtf.caltech.edu/SciUser/.)

One consideration that has been neglected is the possibility that the
planet might show brightness variations with orbital phase.  If a
planet is relatively close to the star, it might be in synchronous
orbit (however the timescale for spin locking the secondary depends
sensitively on $d_P/R_{WD}$ with the 6th power; e.g., see Trilling
2000).  If so, then one would generally expect variable brightness of
the planet as the ``day'' and ``night'' sides alternately face the
Earth during the planet's orbit.  At just a few hundred $R_{WD}$
distant (or tens of $R_J$), the orbital period would be only a few
hours.  However, it is not clear how discrepant the day-night
temperatures will be to yield IR variations.  Changes in the brightness
at UV and optical wavelengths could vary by as much as $\approx 1\pm
A_P(\lambda) R_P^2/ 4d_P^2$ for a suitable viewing perspective, with 
$A_P(\lambda)$ the wavelength dependent reflectivity of the Jovian
atmosphere.  For the classical Roche limit, the minimum orbital
radius will be about $d_P \approx 24 R_J$, so that even if $A_P$
were unity, the maximum brightness variation in reflected light would
be 0.04\%.

It is worth noting that the reflex motion of the star from the presence
of the orbiting planet might be measurable.  Observed Balmer lines tend
to have broad sloping wings but a relatively narrow absorption core.
For example, the core widths in the H$\alpha$ profiles of
Heber, Napiwotzki, \& Reid (1997) and Koester \etal\ (1998) are around
1~\AA, corresponding to velocity widths of 40--50~km s$^{-1}$.  Although the
typical critical speed of rotation for a White Dwarf is quite high
($v_{\rm crit} \approx 4600$~km/s$^{-1}$), the observations are
consistent with little or no rotation (e.g., upper limits of tens of
km s$^{-1}$ from Heber \etal\ 1997 and Koester \etal\ 1998).  For purposes
of a quantitative estimate, consider a White Dwarf of $1M_\odot$ with a
core line width of about 40~km s$^{-1}$, and a planet of $1M_J$ orbiting at
$1000R_{WD}$ with a period of about 3.2~days.  Assuming a circular
orbit, as has been the case throughout this paper, the reflex motion
imparted to the star will be $\approx 150$~m s$^{-1}$.  However, the
problem with White Dwarfs is that they are generally faint.  Of the
local sample, the brightest star is about $m_V=10.5$, and most are
fainter than 12$^{th}$ magnitude.  To detect the reflex motion of such
a speed with 10\% accuracy would require a spectral resolution
$\lambda/\Delta \lambda = 2\times 10^6$ (or at least a centroiding
algorithm to that accuracy), which demands long integration times on
large telescopes for such faint targets.  Note that Jovians at smaller
$d_P$ would produce larger reflex motions in the central star as
$d_P^{-1/2}$; however, the orbital period decreases as $d_P^{3/2}$,
placing an additional constraint on the integration time.

Finally in closing, the SEDs have been derived under the simplifying
assumption of blackbody source functions.  The gas giant would be
expected to show absorption features by molecules, for example, and
these would modify the quantitative predictions of the run of flux
density with frequency for the combined SEDs.  Modification of the
planet temperature by internal heat has also been ignored.  Three of
the gas giants in our solar system show excess IR emission by about
50\% above what is absorbed in sunlight.  Even so, the results
presented in this paper provide quantitative flux density values that
are likely adequate for motivating a search of planets around the
nearest single White Dwarf stars.

\acknowledgements

It is with great pleasure that I thank Steve Spangler for comments
on an early draft of this paper.  I also wish to express appreciation
to an anonymous referee who made several valuable comments.  This
work was supported by a NSF grant (AST-9986915).

\end{document}